\documentclass[onecolumn,a4paper,IEEEtrans]{article}
\usepackage[margin=0.75in]{geometry}
\usepackage{amsmath,graphicx}
\usepackage{mathtools, nccmath} %https://tex.stackexchange.com/questions/433101/rounding-to-nearest-integer-symbol-in-latex
\usepackage{blindtext}
\usepackage{hyperref}
\usepackage{listings}
\usepackage{lipsum}
\newsavebox{\codebox}
\usepackage{tabularx}
\usepackage[table]{xcolor}
\usepackage{colortbl}
\definecolor{lightgreen}{RGB}{200,210,200}
\definecolor{lightred}{RGB}{210,200,200}
\definecolor{lightblue}{RGB}{200,200,210}

\newcommand{\mycc}{\cellcolor{lightgreen}}

\title{Severity classification of ground-glass opacity via \\2-D convolutional neural networks and lung CTs:\\a 3-day exploration}
\date{}
\author{Lisa Y.~W. Tang}
 
\begin{document}
\maketitle
%\ninept

\begin{abstract}
Ground-glass opacity is a hallmark of numerous lung diseases, including patients with COVID19 and pneumonia, pulmonary fibrosis, and tuberculosis. This brief note presents experimental results of a proof-of-concept framework that got implemented and tested over three days as driven by the third challenge entitled "COVID-19 Competition", hosted at the AI-Enabled Medical Image Analysis Workshop of the 2023 IEEE International Conference on Acoustics, Speech and Signal Processing (ICASSP 2023). Using a newly built virtual environment (created on March 17, 2023), we investigated various pre-trained two-dimensional convolutional neural networks (CNN) such as Dense Neural Network, Residual Neural Networks, and Vision Transformers, as well as the extent of fine-tuning each architecture would require. Based on empirical experiments, we opted to fine-tune them using ADAM's optimization algorithm with a standard learning rate of 0.001 for all CNN architectures and apply early-stopping whenever the validation loss reached a plateau. For each trained CNN, the model state with the best validation accuracy achieved during training was stored and later reloaded for new classifications of unseen samples drawn from the validation set provided by the challenge organizers. According to the organizers, few of these 2D CNNs yielded performance comparable to the baseline developed by the organizers. As part of the challenge requirement, the source code produced during the course of this exercise is posted at \url{https://github.com/lisatwyw/cov19}. We also hope that other researchers will find this  prototype of few Python files based on PyTorch 1.13.1 and TorchVision 0.14.1 approachable for their own work.

\end{abstract}
\vspace{1em}

\DeclarePairedDelimiter{\nint}\lfloor\rceil
\DeclarePairedDelimiter{\abs}\lvert\rvert %https://tex.stackexchange.com/questions/433101/rounding-to-nearest-integer-symbol-in-latex

\section{Introduction}
\label{sec:intro}

Ground-glass opacity (GGO) as captured in lung computed tomography scans (CTs) is a hallmark of numerous lung
diseases and typically signifies lung consolidation \cite{engeler93,ggo}. The extent of GGO observed in lung scans is one approach for severity assessment (known as `grading') that has been used in clinical practice to facilitate the triage of patients in hospitals and acute care centres, albeit the manual process of grading by visual inspection of lung scans is laborious and time-consuming. 

With the advent of deep convolutional neural networks (CNNs), computer-automated and/or computer-assisted classification of image volumes may now be done with high efficiency and accuracy. %numerous studies have shown the promise of computer-assisted severity . %Thanks to the plethora of open-source frameworks and software packages, deep and wide convolutional neural networks (CNNs) may be trained with and without fine-tuning, thus rendering this overly ambitious effort feasible. 
Countless studies from the past five years, e.g. \cite{tang2020,shaik2022transfer}, have further shown the success of transfer learning of CNNs trained using three-channel two-dimensional inputs. Accordingly, we approach severity classification of GGO similarly to leverage off-the-shelf pretrained networks that could facilitate model training with training sample sizes smaller than 500. To this end, we explored the use of various CNNs architectures: AlexNet \cite{AlexNet}, VGG \cite{VGG}, Residual Networks \cite{ResNet},  Wide Residual Networks \cite{WideResNetNet},  DenseNet \cite{DenseNet}, SqueezeNet \cite{SqueezeNet} and Vision Transformers \cite{VTB32}.

This note aims to document the initial experimental work done on the open-source dataset named ``COV19CT-DB'' (COVID-19 Computed Tomography Database) as part of a challenge submission to the third competition entitled  ``ICASSP: COVID19 severity detection challenge''. Due to time constraints (i.e. three days), we focus on the severity classification problem where the objective is to label each CT volume as either mild, moderate, severe, or critical \cite{kollias2022ai}. These categories were predetermined by a pre-existing protocol where experts visually inspected each scan for GGO and manually labeled each as ``mild'' if less than 26\% of the lung volumes were judged to capture ``pulmonary parenchymal involvement'';  ``moderate'' if the involvement as 26-50\%; ``severe'' if involvement is 50-70\%; or ``critical'' if involvement was greater than 75\% \cite{morozov2020mosmeddata}. 

\section{Materials}

The training and validation samples is derived as a subsest of a larger dataset named ``COV19-CT-DB'' that contains over one thousand patient samples of three-Dimensional CT chest scans as archived in lossy compression format (i.e. JPEG) \cite{kollias2022ai,arsenos2022large,kollias2021mia,kollias2020deep,kollias2020transparent,kollias2018deep}. %As further elaborated in previous articles, 
These CT volumes were collected from different hospitals in the United States from September 1 2020 and November 30
2021 \cite{kollias2022ai}. Upon successful enrolment into the ICASSP 2023 challenge, participants were directed to four hyperlinks to OneDrive where zipped archives (.rar or .zip) could be downloaded. Upon decompression, each folder represents a patient scan, with each containing up to 300 individual CT slices in JPEG format. To this end, 430 and 101 CT scans were provided for model training and validation, respectively.

\iffalse
import nrrd

# Some sample numpy data
data = np.zeros((5,4,3,2))
filename = 'testdata.nrrd'

# Write to a NRRD file
nrrd.write(filename, data)

# Read the data back from file
readdata, header = nrrd.read(filename)
print(readdata.shape)

\fi 

\section{Methods}
\label{sec:methods}

On a high level, we propose the deployment of pretrained convolutional neural networks (CNNs) for the classification task.  
For ease of optimization, our approach explores a previous framework that leveraged two-dimensional (2D)  Residual Networks \cite{tang2020} for the identification of a lung disease. The decision to employ 2D networks was motivated by two factors: time constraints and the observation that GGO have been observed in lower lung lobes in  COVID-19 patients \cite{kollias2022ai}.

\subsection{Preprocessing}

As the information on field of volume and voxel spacing are no longer accessible in the archival format provided by the challenge (i.e. JPEG), the  depth  of each CT volume was approximated by the file count $n$ of each scan / folder, i.e. 

\begin{lrbox}{\codebox}
\begin{lstlisting}[language=python,basicstyle=\small]
import subprocess
output = subprocess.run(['ls', ct_folder], stdout = subprocess.PIPE ); 
n = len( output.stdout.decode('utf-8').split() )    
\end{lstlisting}
\end{lrbox}

\noindent 
\colorbox{gray!10}{
\setlength{\fboxrule}{0ex} % required
\fbox{%
  \begin{minipage}{\dimexpr\linewidth-0ex}
  \usebox{\codebox}    
  \end{minipage}%
  }
}

The axial section centered at $z$ was computed deterministically ($z=\nint{n*f}$). The value of $f=0.25$ was defined based on visual inspection on a subset of the images randomly drawn from the training set. Then, three contiguous slices centered at $z$ were concatenated to form three-channel inputs of size $3 \times v \times v$ where $v$ depends on the CNN model \cite{tang2020}.

Prior to resizing, lung masks were also computed to mask out regions outside the rib cage, as shown in Figures 1-3. The implementation code of lung mask generation was adopted from an online resource \cite{kaggle} such that the threshold parameter  is adjusted so it will operate on the compressed data as stored in the JPEG images from COV19-CT-DB, i.e.:

Figures 1-3 provide examples of the image inputs.

\begin{lrbox}{\codebox}
\begin{lstlisting}[language=python,basicstyle=\small]
from skimage.segmentation import clear_border
init_roi_mask = jpeg_im_slice<100 # Updated 
roi_mask_v0 = clear_border(init_roi_mask)
...
\end{lstlisting}
\end{lrbox}

\noindent
\colorbox{gray!10}{
\setlength{\fboxrule}{0ex}
    \fbox{%
      \begin{minipage}{\dimexpr\linewidth-0ex}
      \usebox{\codebox}    
      \end{minipage}%
    }
}

%[language=pascale,numbers=left, numberstyle=\small, stepnumber=2, numbersep=5pt]

%the voxel intensity values were clipped using a window/level of 350 Hounsfield units (HU)/−1150 HU and normalized to the range of [0, 1]

\subsection{Model-training}

Each model architecture was fine-tuned over a maximum of 500 epochs. We used the categorical cross-entropy objective.
For all CNN architectures, we applied early-stopping whenever the validation loss reached a plateau. Two optimization algorithms explored were Adaptive Moment Estimation (ADAM) and Stochastic Gradient Descent (SGD). For SGD, the standard setting of using momentum value of 0.9 was used. The following settings were explored initially: batch size BS=$\{16, 128\}$, optimization's learning rate LR=$\{0.001, 0.01\}$. %Further, to enhance robustness against deformations and scale transformations, we explored the application of random  horizontal flips and random rigid transformations, applied at 50\% of the times (probability of 0.5). 

\subsection{Extent of fine-tuning}
We initialized each CNN with pretrained weights and subsequently explored two level of network fine-tuning: allowing the network weights of all layers or only the last layer to be changed/optimized.  

\subsection{Evaluation metrics and model selection}

For each class, we calculated the area under the receiver's operating curve (AUROC) and the F1-macro score, i.e.
\begin{equation}
\textbf{F1-mac} =  \frac{1}{K} \sum^K_{i=1}  \frac{2 \times PR_i \times RC_i }{ PR_i + RC_i }    
\end{equation}
\noindent where $K$ is the number of classes ($K=4$ severity classes of GGO) and $RC_i$ and $PR_i$ respectively denotes the precision and recall for class $i$, i.e. $PR_i = \frac{TP_i}{TP_i+FP_i}$ and $RC_i =\frac{TP_i}{ TP_i+FN_i }$. 

For each trained CNN, the state with the best validation accuracy achieved during training was selected for the evaluation of each test sample. 

\subsection{Evaluation protocol}

We set a random subset of the training set as the internal validation set. The entire validation set of $n$=101 was left as unseen by each CNN. The test set ($n$=230) was provided by the organizers a few days before the challenge deadline. The labels on the test set is blinded to participants.

\begin{table}[ht]
\label{tab:res1}
\begin{center}\small
\begin{tabular}{l | l | l  l  | l  l  | l } 
 \hline
 &  & \multicolumn{2}{c|}{AUROC}  &\multicolumn{2}{c|}{ F1-macro} &  Pred. class distr. \\ 
Model & Settings& Val & Unseen*  & Val & Unseen*  \\ \hline
AlexNet &  BS16 SGD LR0.001 & 66.8 & 59.0 & 49.6 & 39.5 & 89, 31, 92, 19 \\ \hline
AlexNet &  BS32 ADAM LR0.001 & 66.6 & 59.8 & 51.1 & 34.0 & 61, 53, 106, 11 \\ \hline
AlexNet &  BS512 SGD LR0.001 & 67.1 & 59.0 & 39.9 & 30.9 &  93, 18, 120, 0 \\ \hline
AlexNet &  BS64 ADAM LR0.001 & 66.1 & 62.0 & 49.4 & 45.1 &  78, 41, 100, 12 \\ \hline
DenseNet201 &  BS16 SGD LR0.001 & 62.1 & 59.0 & 42.9 & 38.8 &  80, 32, 114, 5 \\ \hline
DenseNet201 &  BS32 ADAM LR0.001 & 63.4 & 63.3 & 46.3 & 46.8 & 57, 73, 99, 2 \\ \hline
DenseNet201 &  BS512 SGD LR0.001 & 63.1 & 62.5 & 45.6 & 45.1 &  50, 76, 103, 2 \\ \hline
DenseNet201 &  BS64 ADAM LR0.001 & 61.4 & 62.2 & 41.9 & 36.4 &  57, 63, 109, 2 \\ \hline
DenseNet &  BS16 ADAM LR0.001 & 63.8 & 60.1 & 34.8 & 35.9 &  86, 15, 130, 0 \\ \hline
DenseNet &  BS16 SGD LR0.001 & 63.8 & 60.1 & 34.8 & 35.9 &  86, 15, 130, 0 \\ \hline
DenseNet &  BS32 ADAM LR0.001 & 65.1 & 67.2 & 38.1 & 40.9 &  71, 26, 134, 0 \\ \hline
DenseNet &  BS512 SGD LR0.001 & 67.5 & 67.0 & 41.5 & \mycc 51.4 & 68, 46, 115, 2 \\ \hline
DenseNet &  BS64 ADAM LR0.001 & 65.8 & 63.4 & 39.2 & 37.6 &  58, 17, 156, 0 \\ \hline
InceptionNet &  BS16 SGD LR0.001 & 59.5 & 58.2 & 38.1 & 40.0 &  89, 17, 123, 2 \\ \hline
InceptionNet &  BS32 ADAM LR0.001 & 63.0 & 52.5 & 44.0 & 27.8 &  80, 41, 108, 2 \\ \hline
InceptionNet &  BS512 SGD LR0.001 & 63.7 & 64.3 & 38.6 & 39.5 &  39, 58, 134, 0 \\ \hline
InceptionNet &  BS64 ADAM LR0.001 & 60.8 & 55.1 & 40.7 & 36.3 & 61, 77, 91, 2 \\ \hline
ResNet152 &  BS16 SGD LR0.001 & 63.8 & 54.6 & 36.5 & 28.5 &  76, 51, 104, 0 \\ \hline
ResNet152 &  BS32 ADAM LR0.001 & 67.7 & 57.5 & 40.7 & 29.5 &  74, 35, 122, 0 \\ \hline
ResNet152 &  BS512 SGD LR0.001 & 66.6 & 59.5 & 39.4 & 30.8 & 107, 50, 74, 0 \\ \hline
ResNet152 &  BS64 ADAM LR0.001 & 65.1 & 56.3 & 38.4 & 28.8 &  81, 26, 124, 0 \\ \hline
SqueezeNet &  BS16 SGD LR0.001 & 68.8 & 58.8 & 42.6 & 32.4 &  74, 34, 123, 0 \\ \hline
SqueezeNet &  BS32 ADAM LR0.001 & 67.7 & 63.4 & 53.2 & 44.6 & 61, 68, 95, 7 \\ \hline
SqueezeNet &  BS512 SGD LR0.001 & 65.4 & 58.9 & 50.3 & 39.0 &  60, 54, 112, 5 \\ \hline
SqueezeNet &  BS64 ADAM LR0.001 & 66.5 & 63.8 & 50.5 & 45.2 & 60, 70, 94, 7 \\ \hline
VGG &  BS16 SGD LR0.001 & 62.5 & 58.5 & 43.9 & 35.4 & 75, 65, 81, 10 \\ \hline
VGG &  BS32 ADAM LR0.001 & 62.2 & 63.7 & 45.1 & 44.2 & 89, 65, 72, 5 \\ \hline
VGG &  BS512 SGD LR0.001 & 63.1 & 60.5 & 44.6 & 33.5 & 66, 78, 82, 5 \\ \hline
VGG &  BS64 ADAM LR0.001 & 63.6 & 58.2 & 46.8 & 33.0 & 92, 57, 78, 4 \\ \hline
VTB32 &  BS16 SGD LR0.001 & 68.6 & 68.0 & 54.7 & \mycc 52.0 & 77, 45, 95, 14 \\ \hline
VTB32 &  BS32 ADAM LR0.001 & 67.1 & 65.7 & 52.9 & 48.9 & 78, 48, 96, 9 \\ \hline
VTB32 &  BS512 SGD LR0.001 & 63.6 & 61.8 & 45.2 & 40.4 &  78, 38, 115, 0 \\ \hline
VTB32 &  BS64 ADAM LR0.001 & 67.0 & 66.6 & 52.6 & \mycc 51.3 & 74, 48, 96, 13 \\ \hline
WideResNet101 &  BS16 SGD LR0.001 & 67.0 & 60.2 & 40.4 & 33.4 &  84, 27, 120, 0 \\ \hline
WideResNet101 &  BS32 ADAM LR0.001 & 65.0 & 58.3 & 37.4 & 29.9 & 90, 16, 125, 0 \\ \hline
WideResNet101 &  BS512 SGD LR0.001 & 65.3 & 57.2 & 48.4 & 31.5 & 73, 55, 101, 2 \\ \hline
WideResNet101 &  BS64 ADAM LR0.001 & 62.1 & 58.1 & 32.6 & 29.7 & 89, 6, 136, 0 \\ \hline
\end{tabular}
\caption{Performance when model weights were fine-tuned \textbf{only on the last layer}. Shown are the overall and class-wise F1-scores expressed in percentages. *Performance of CNNs when evaluated the entire validation set was not observed during training. ``Pred. class distr.'' denotes the predicted class distribution on the test set.} %``FT'' denotes the extent of fine-tuning: network weights from all layers or only the last layer were optimized. 
\end{center}
\end{table}

\begin{table}[ht]
\label{tab:res2}
\begin{center}\small
\begin{tabular}{l | l | l  l  | l  l  | l } 
 \hline
& & \multicolumn{2}{c|}{AUROC}  &\multicolumn{2}{c|}{ F1-macro} &  Pred. class distr. \\ 
Model & Settings & Val & Unseen  & Val & Unseen  \\ \hline 
AlexNet &  BS16 SGD LR0.001 & 50.0 & 50.0 & 13.8 & 15.4 &  0, 0, 231, 0 \\ \hline
AlexNet &  BS32 ADAM LR0.001 & 50.0 & 50.0 & 13.8 & 15.4 &  0, 0, 231, 0 \\ \hline
AlexNet &  BS512 SGD LR0.001 & 42.6 & 46.7 & 23.8 & 21.2 &  0, 25, 206, 0 \\ \hline
AlexNet &  BS64 ADAM LR0.001 & 50.0 & 50.0 & 13.8 & 15.4 &  0, 0, 231, 0 \\ \hline
DenseNet201 &  BS16 SGD LR0.001 & 68.3 & 63.4 & 52.9 & 43.1 & 101, 21, 95, 14 \\ \hline
DenseNet201 &  BS32 ADAM LR0.001 & 72.1 & 63.1 & 59.4 & 43.7 & 71, 75, 74, 11 \\ \hline
DenseNet201 &  BS64 ADAM LR0.001 & 72.2 & 66.4 & 59.3 & 49.4 & 59, 81, 85, 6 \\ \hline
DenseNet &  BS16 ADAM LR0.001 & 68.2 & 69.7 & 54.3 & \mycc 51.6 & 81, 64, 79, 7 \\ \hline
DenseNet &  BS16 SGD LR0.001 & 67.9 & 70.4 & 53.6 & \mycc 54.1 &  58, 43, 125, 5 \\ \hline
DenseNet &  BS32 ADAM LR0.001 & 74.7 & 65.7 & 63.5 & 47.6 & 84, 81, 58, 8 \\ \hline
DenseNet &  BS64 ADAM LR0.001 & 71.1 & 64.3 & 54.8 & 41.8 & 105, 15, 87, 24 \\ \hline
InceptionNet &  BS16 SGD LR0.001 & 68.3 & 61.5 & 51.8 & 38.4 & 107, 56, 63, 5 \\ \hline
InceptionNet &  BS32 ADAM LR0.001 & 73.1 & 66.4 & 60.4 & 48.2 & 86, 52, 79, 14 \\ \hline
InceptionNet &  BS64 ADAM LR0.001 & 66.8 & 58.6 & 45.6 & 32.2 & 115, 4, 108, 4 \\ \hline
ResNet152 &  BS16 ADAM LR0.001 & 72.5 & 71.9 & 56.7 & \mycc 57.3 & 94, 65, 47, 25 \\ \hline
ResNet152 &  BS16 SGD LR0.001 & 67.2 & 68.4 & 47.5 & 48.7 & 112, 70, 40, 9 \\ \hline
ResNet152 &  BS32 ADAM LR0.001 & 68.4 & 65.1 & 52.5 & 47.5 & 87, 51, 88, 5 \\ \hline
ResNet152 &  BS64 ADAM LR0.001 & 67.6 & 71.7 & 49.4 & \mycc 56.2 & 85, 77, 65, 4 \\ \hline
SqueezeNet &  BS16 SGD LR0.001 & 59.3 & 62.8 & 32.2 & 37.0 &  48, 13, 170, 0 \\ \hline
SqueezeNet &  BS32 ADAM LR0.001 & 63.7 & 61.1 & 38.0 & 34.4 & 61, 36, 134, 0 \\ \hline
SqueezeNet &  BS512 SGD LR0.001 & 67.5 & 65.3 & 41.9 & 39.2 &  61, 33, 137, 0 \\ \hline
SqueezeNet &  BS64 ADAM LR0.001 & 68.8 & 67.5 & 41.4 & 40.1 & 111, 34, 86, 0 \\ \hline
VGG &  BS16 SGD LR0.001 & 69.8 & 59.8 & 53.8 & 36.8 & 88, 46, 71, 26 \\ \hline
VGG &  BS32 ADAM LR0.001 & 63.7 & 58.5 & 44.9 & 34.4 & 76, 50, 86, 19 \\ \hline
VGG &  BS64 ADAM LR0.001 & 70.1 & 68.7 & 54.9 & 49.0 & 116, 64, 35, 16 \\ \hline
VTB32 &  BS16 SGD LR0.001 & 62.1 & 64.1 & 42.4 & 46.0 & 84, 55, 89, 3 \\ \hline
VTB32 &  BS32 ADAM LR0.001 & 60.8 & 61.8 & 39.2 & 40.1 & 103, 5, 116, 7 \\ \hline
VTB32 &  BS512 SGD LR0.001 & 67.7 & 61.9 & 53.4 & 44.6 & 49, 64, 115, 3 \\ \hline
VTB32 &  BS64 ADAM LR0.001 & 69.1 & 67.4 & 54.2 & \mycc 52.1 & 68, 43, 109, 11 \\ \hline
WideResNet101 &  BS16 SGD LR0.001 & 69.6 & 62.2 & 54.3 & 42.0 & 120, 34, 69, 8 \\ \hline
WideResNet101 &  BS32 ADAM LR0.001 & 68.1 & 65.9 & 52.2 & 49.5 & 33, 72, 118, 8 \\ \hline
WideResNet101 &  BS64 ADAM LR0.001 & 68.4 & 63.1 & 50.0 & 44.6 & 80, 90, 57, 4 \\ \hline
\end{tabular}
\caption{Performance when \textbf{all} model weights were fine-tuned.}
\end{center}
\end{table}

\section{Results}

We first examined the results of fine-tuning the models on the last layer. Table 1 reports preliminary comparisons of the accuracies (F1-macro score and area under receiver's operating curve) achieved by individual models as evaluated on the unseen validation set. That is, all images from the validation set were not observed by the models during training. As Table 1 shows, only DenseNet and Vision Transformer (VTB32) achieved greater than 51 in F1-macro on the unseen validation set ($n$=101). 

We next examined the results of fine-tuning the models on the all layers. Similar to the previous results, only DenseNet, ResNet152, and Vision Transformer (VTB32) achieved greater than 51 in F1-macro on the same unseen validation set, as reported in Table 2. The class-wise F1-macro scores are further reported in Table 3. 

In summary, when only the training set was used ($n$=430) and the entire validation set was left as unseen (i.e. only used for evaluation), fine-tuning all network weights seemed to add minor improvement to the accuracy performance of DenseNet and Vision Transformers while fine-tuning ResNet152 substantially improved accuracy (F1-macro of less than 31.0 increased to greater than 56.0).

 %More specifically, ResNet152 achieved mean F1-macro of 56.2 and classwise F1-macro scores of 100.0, 79.4, 68.8, and 71.7. The second top performing model achieved mean F1-macro of 56.2 and classwise F1-macro scores of .

Based on these empirical results, we retrained ResNet152 using the same settings as listed in Table 3 but using both the training and validation images for model training.
\\
\\
\textbf{NB}. This section will be updated when the third-party evaluations of the two prediction files submitted to the organizers are published. 

\begin{table}[ht]
\label{tab:summary}
\begin{center}\small
\begin{tabular}{l l | l | l l l l } 
 \hline
 & & &\multicolumn{4}{c}{Class-wise F1-macro}   \\ 
Model & Settings & Average F1-macro  & Mild & Moderate & Severe & Critical \\ \hline 
DenseNet & BS16 SGD LR0.001 & 54.1 & 100 & 74.4 & 67.6 & 62.8 \\
ResNet152 & BS16 ADAM LR0.001  & 56.2 & 100 & 79.4& 68.8&  71.7 \\ 
VT (32-bit) & BS64 ADAM LR0.001 &  52.1 & 100& 74.7 &70.6 &65.0 \\ \hline 
\end{tabular}
\caption{Summary of performance metrics reported when weights of \textbf{all} layers were fine-tuned.}
\end{center}
\end{table}

\begin{figure*}[t]\centering
\includegraphics[width=1\textwidth]{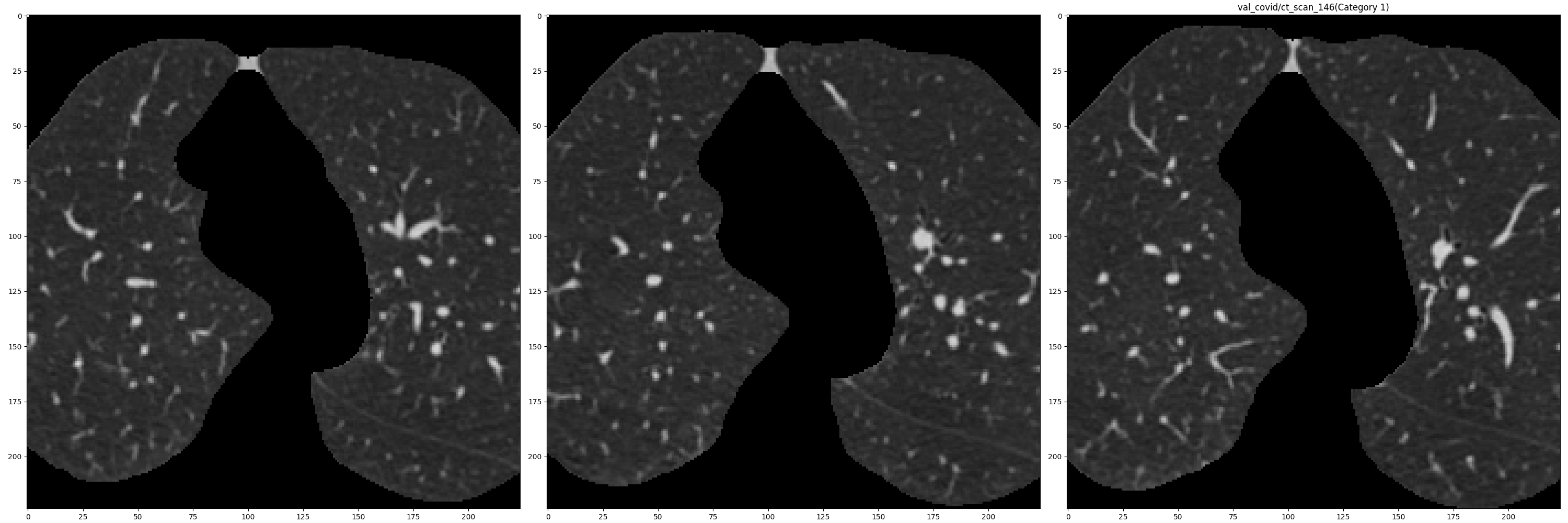}
\caption{Example training input used to fine-tune CNNs.}
\label{fig:in}
\end{figure*}

\begin{figure*}[t]\centering
\includegraphics[width=1\textwidth]{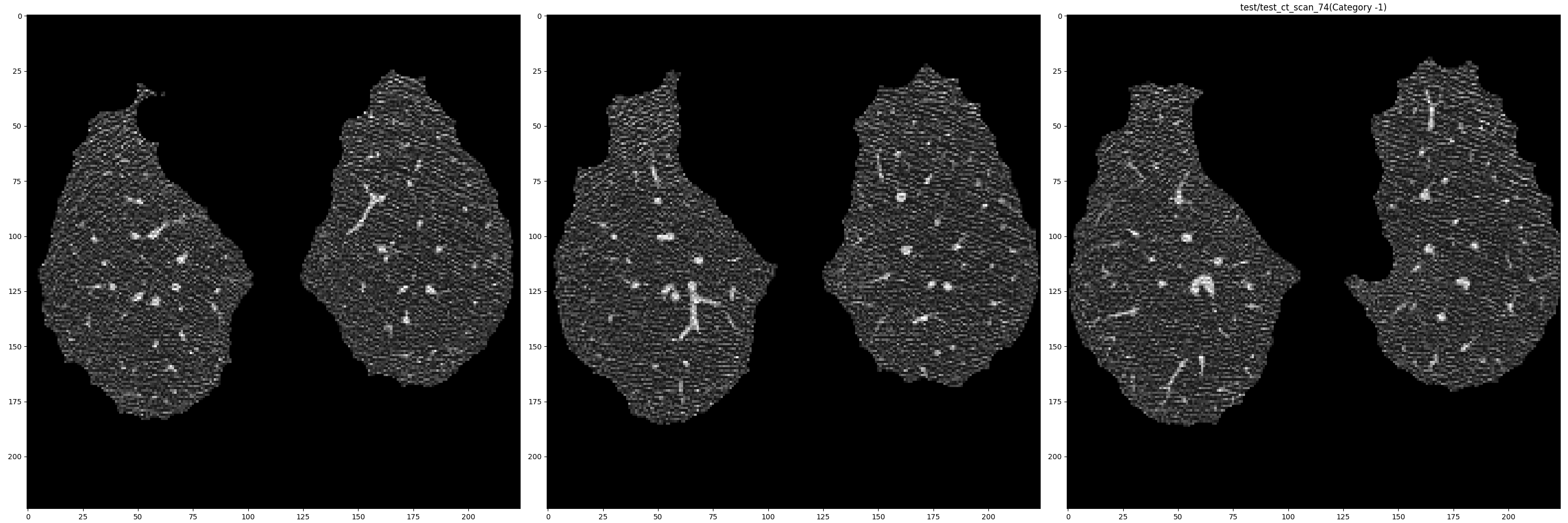}
\caption{Example test input.}
\label{fig:in2}
\end{figure*}

\begin{figure*}[t]\centering
\includegraphics[width=1\textwidth]{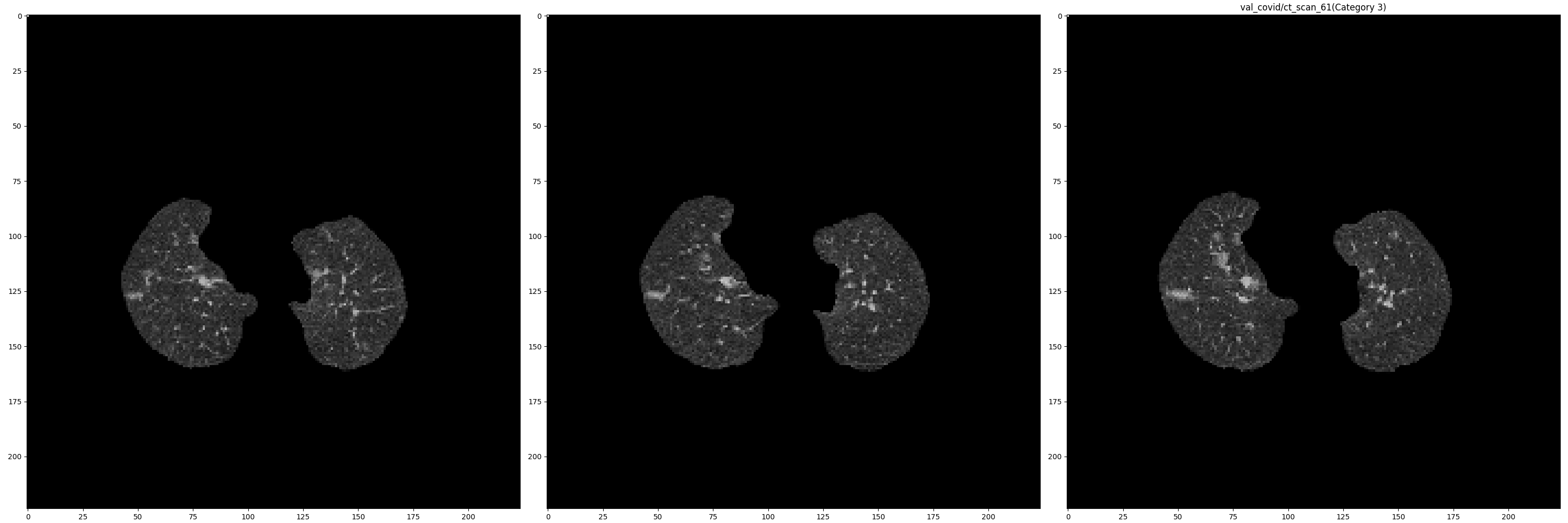}
\caption{Another training input drawn from the severe class without application of the center-cropping.}
\label{fig:in3}
\end{figure*}

\section{Conclusion}

In this brief note, we shared empirical data that explored the feasibility of severity classification without the deployment of three-dimensional neural networks. 
The source code developed during the course of this experimental prototyping period is posted at \url{https://github.com/lisatwyw/cov19}. We hope that other researchers may find this quick prototype consisting of few Python files based on
PyTorch 1.13.1 and TorchVision 0.14.1 approachable.
\\
\newpage
\noindent\textbf{\large{Acknowledgements}}
\\
\\
The author sincerely thank Professor Dimitrios Kollias and the organizing committee for provisioning the COV19-CT-DB dataset and hosting this exciting challenge \cite{kollias2022ai,arsenos2022large,kollias2021mia,kollias2020deep,kollias2020transparent,kollias2018deep}. The author also expresses deep gratitude to Tong Tsui Shan and Kim Chuen Tang as well as staff of Compute Canada/Alliance Canada and Data Science Institute for their support.

\appendix

\section{ Efforts to increase transparency and  reproducibility}  

One huge bottleneck the author experienced was the file transfer process, which involved downloading compressed archives from OneDrive onto the author's local computer and uploading them to a remote computing cluster (i.e. AllianceCan). Further, some of the subfolders were archived in .RAR format, which could not be unzipped on the remote cluster. As a workaround, the files were uncompressed on the author's local computer and then subsequently uploaded to the remote cluster. These file transfers may risk lost of data files. To assist future users with ensuring data integrity of the downloaded scans, a Google sheet listing the file counts for each scan has been compiled at this link:
\url{https://docs.google.com/spreadsheets/d/1SoVfioBKj_ElEETEk7o7KK_vs6VEca8LLIYW0xXpSYY/}
\\
\\

\end{document}